\documentclass[11pt,
showpacs,
preprintnumbers,amsmath,amssymb,
aps,prd,nofootinbib,eqsecnum,a4paper]{revtex4}
\usepackage{epsfig}
\usepackage[autostyle]{csquotes}
\usepackage{graphicx,epsf}
\usepackage{color}
\usepackage{dirtytalk}
\usepackage{bm}
\usepackage{psfrag}
\usepackage[symbol]{footmisc}
\usepackage{tensor}
\usepackage{hyperref}

\def\beq{\begin{equation}}
\def\eeq{\end{equation}}
\def\bea{\begin{eqnarray}}
\def\eea{\end{eqnarray}}
\def\bi{\begin{itemize}}
\def\ei{\end{itemize}}
\def\cs2{c_{\rm{s}}^2}
\def \beg {\begin{enumerate}}
\def \en {\end{enumerate}}

\def\M0{{\cal M}_0}

\begin{document}

\title{A comparison between the Jordan and Einstein frames in Brans-Dicke theories with torsion}

\author{R. Gonzalez Quaglia\footnote[1]{\href{mailto:rodrigo@icf.unam.mx}{rodrigo@icf.unam.mx}}}
\author{Gabriel Germ\'an\footnote[2]{\href{mailto:gabriel@icf.unam.mx}{gabriel@icf.unam.mx}}}
\affiliation{
Instituto de Ciencias F\'{i}sicas, Universidad Nacional Aut\'{o}noma de M\'{e}xico,\\Av. Universidad S/N. Cuernavaca, Morelos, 62251, M\'{e}xico}
%

\begin{abstract}
In recent years, gravitational models motivated by quantum corrections to gravity which introduce higher order terms like $R^{2}$ or terms in which the Riemann tensor is not symmetric have been studied by several authors in the form of a general Brans-Dicke type model containing the Ricci scalar, the Holst term and the Nieh-Yan invariant. In this paper we focus on the less  explored Jordan frame of such theories and in the comparison between both this frame and the Einstein one.  Furthermore, we discuss the role of the transformation of the torsion under conformal transformations and show that the transformation proposed in this paper (extended conformal transformation) contains a special case of the projective transformation of the connection used in some of the papers that motivated this work. We discuss the role and advantages of the extended conformal transformation and show that this new approach can have interesting consequences by working with different variables such as the metric and torsion. Moreover, we study the stability of the system via a dynamical analysis in the Jordan frame, this in order to analyze whether or not we have the fixed points that can be later identified as the inflationary attractor and the unstable fixed point where inflation could take place. Finally we study the scale invariant case of the general model in the Jordan frame. We  find out that both the scalar spectral index and the tensor-to-scalar ratio are in agreement with the latest Planck results. 
  
\end{abstract}


\maketitle

\section {\bf Introduction}\label{sec:intro}
\subsection{Addition of torsion.}Usually, when we think about a theory of gravitation, General Relativity (GR) comes to mind \cite{The Foundation of the 
General Theory of Relativity}. GR is a geometric theory which uses the curvature of space-time to fully describe a manifold. However, the curvature is not the only object than can fully describe a manifold and thus GR does not have a unique formulation, there exists at least two different ones that are classically equivalent to GR. One of these formulations, teleparallel gravity has been widely studied not only by Einstein himself \cite{Einheitliche Feldtheorie von Gravitation und Elektrizitat}-\cite{ Riemann-Geometrie unter Aufrechterhaltung des Begriffes des Fernparallelismus} but by many more recent authors such as \cite{Teleparallel gravity}-\cite{Higgs inflation and teleparallel gravity}. A second formulation is the so called symmetric teleparallel  formulation \cite{Symmetric teleparallel general relativity}. Apart from these formulations, there exists at least two variational principles that give rise to the same field equations (starting from the Einstein-Hilbert action). Within this work we are specially interested in the Palatini formalism (also found by Einstein) \cite{Palatini} which consists on considering the connection $\Gamma^{\lambda}_{\mu\nu}$ and the metric tensor $g_{\mu\nu}$ as two separate objects that are independent of one another. As already mentioned, the Palatini variation is classically equivalent to GR but if we work on a modified version or extension of GR, this equivalence is not, in general,  true anymore. 
Another key assumption within GR is the fact that the metric has vanishing covariant derivative,  the so called \enquote{non-metricity tensor} is null.
\begin{equation}
Q_{\alpha\mu\nu}=\nabla_{\alpha}g_{\mu\nu}=0, 
\end{equation}
and that the connection is symmetric on the two lower indices 
\begin{equation}
    \Gamma^{\lambda}_{\mu\nu}=\Gamma^{\lambda}_{\nu\mu}.
\end{equation}
However, until today, these two assumptions have no good motivation. Indeed one can relax these two assumptions and introduce non zero torsion and non-metricity as building blocks of a theory of gravity. In fact this idea leads to the previously mentioned formulations of GR, the teleparallel formulation uses torsion as its main object and the symmetric teleparallel formulation uses non-metricity. In this work we will focus on a theory of gravity containing curvature, characterized by the Levi-Civita connection $\mathring{\Gamma}^{\lambda}_{\mu\nu}$ and a non symmetric connection that gives rise to torsion which is defined as 
\begin{equation}
T^{\lambda}_{\mu\nu} =\Gamma^{\lambda}_{\mu\nu}-\Gamma^{\lambda}_{\nu\mu}.
\end{equation}
We will consider the model presented in \cite{Langvik} and \cite{Higgs inflation in Einstein-Cartan gravity} in which the torsion is different from zero but with still vanishing non-metricity. First we will briefly explain this model and its features for us to later be able to compare the results in these two references with the ones found in this paper.
Having a non zero torsion tensor will define a general connection that includes the Levi-Civita one plus a contribution coming from the torsion, this general connection is defined as 
\begin{equation}\label{connection}
    \Gamma^{\lambda}_{\ \mu\nu}=\mathring{\Gamma}^{\lambda}_{\ \mu\nu}+K^{\lambda}_{\ \mu\nu}.
\end{equation}
Here, the contorsion tensor $K^{\lambda}_{\ \mu\nu}$ is defined as 
\begin{equation}
K_{\alpha\beta\gamma}=\frac{1}{2}\left(T_{\alpha\beta\gamma}+T_{\gamma\alpha\beta}+T_{\beta\alpha\gamma}\right).
\end{equation}
The Riemann tensor $R^{\lambda}_{\ \mu\nu\rho}$ is now constructed with the connection (\ref{connection})
\begin{equation}
R^{\lambda}_{\ \mu\nu\rho}=\Gamma^{\lambda}_{\ \rho\mu,\nu}-\Gamma^{\lambda}_{\ \nu\mu,\rho}+\Gamma^{\lambda}_{\ \nu\sigma}\Gamma^{\sigma}_{\ \rho\mu}-\Gamma^{\lambda}_{\ \rho\sigma}\Gamma^{\sigma}_{\ \nu\mu}.
\end{equation}
This will give rise to the usual Riemann tensor used in GR and a new contribution coming from the torsion part of the connection. Moreover, as the connection is no longer symmetric, also the Riemann tensor losses its symmetry giving rise to new terms that can be considered in a gravitational action. 

\subsubsection{Holst term and Nieh-Yan invariant.} In GR we can only built one scalar that is quadratic on derivatives of the metric from contractions of the Riemann tensor, this is the so called Ricci scalar defined as 
\begin{equation}\label{RiciScalar}
R=g^{\alpha\beta}R^{\mu}_{\ \alpha\mu\beta}.
\end{equation}
However, now we can also construct another scalar by contracting the Riemann tensor with the  Levi-Civita tensor as 
\begin{equation}\label{Holstterm}
 \mathcal{R}=\frac{1}{2}\epsilon^{\alpha\beta\gamma\delta}R_{\alpha\beta\gamma\delta}.
\end{equation}
This new scalar is known as the Holst term \cite{PARITY VIOLATION IN METRIC TORSION THEORIES OF GRAVITATION}-\cite{Barberos Hamiltonian derived from a generalized Hilbert-Palatini action} that together with the Ricci scalar form the only two scalars that are quadratic on derivatives of the metric. 
The exact form of these two scalars reads 
\begin{equation}\label{RicciTensor}
R=\mathring{R}+T+2\mathring{\nabla}_{\mu}T^{\mu},
\end{equation}
\begin{equation}\label{HolstTerm}
\mathcal{R}=-3\mathring{\nabla}_{\mu}\hat{T}^{\mu}+\frac{1}{4}\epsilon^{\alpha\beta\gamma\delta}T_{\mu\alpha\beta}T^{\mu}_{\ \gamma\delta}.
\end{equation}
Here we used the three contractions of the torsion 
\begin{equation}\label{TorsionVectors}
T^{\mu}=g_{\rho\sigma}T^{\rho\mu\sigma}, \quad \hat{T}^{\mu}=\frac{1}{6}\epsilon^{\mu\nu\rho\sigma}T_{\nu\rho\sigma},
\end{equation}
\begin{equation}\label{TorsionScalar}
T=\frac{1}{4}T_{\mu\nu\rho}T^{\mu\nu\rho}-\frac{1}{2}T_{\mu\nu\rho}T^{\rho\mu\nu}-T_{\mu}T^{\mu}.
\end{equation}
We can now consider that our gravitational theory explicitly contains torsion, therefore we can propose a term like 
\begin{equation}\label{Nieh-Yan}
\partial_{\mu}\epsilon^{\mu\nu\rho\sigma}T_{\nu\rho\sigma},
\end{equation}
to be included in our theory. This is the so called \enquote{Nieh-Yan invariant} \cite{An Identity in Riemann-cartan Geometry}. Note that the torsion scalar is already included in the Ricci scalar.
Hence, our gravitational action takes the following form 
\begin{equation}\label{GravitationalAction}
    S_{g}=\int d^{4}x\sqrt{-g}\left[\frac{1}{2}R+\frac{1}{2}\mathcal{R}+\frac{3}{2}\partial_{\mu}\epsilon^{\mu\nu\rho\sigma}T_{\nu\rho\sigma}\right].
\end{equation}
Note that the matter part of the model will be included latter in the form of a scalar sector, we will neglect any fermionic contribution and we will work in units where the Planck mass $M_{p}=1$ unless otherwise specified. The main objective of this paper is to compare the yet unexplored Jordan frame of the model in \cite{Langvik} and \cite{Higgs inflation in Einstein-Cartan gravity} in order to clarify differences and similarities of the two mentioned frames. This type of comparison have been already carried out in several models and contexts such as in \cite{Einstein frame or Jordan frame?}-\cite{Dark energy: The Equation of state description versus scalar-tensor or modified gravity}. As there is still an ongoing discussion on which of the two frames is the physical one and whether or not they are equivalent to one another, the intended contribution of this paper is to compare these two frames in an explicit model. In \cite{Langvik} and \cite{Higgs inflation in Einstein-Cartan gravity} the authors focused on the analysis of the Einstein frame of the model (\ref{GenericAction}), rather, in the present paper we will focus on the Jordan frame of the cited model and in the comparison of the results found in each frame. In the last section of the paper we study the inflationary solutions of the scale invariant version of our generic model. Here the torsion terms account only as a modification of the kinetic term of the scalar field following the procedure shown in  Sections \ref{The model}, \ref{Einstein Frame} and \ref{Jordan Frame}. Other works involving torsion and cosmology can be found in \cite{Cosmic relic torsion from inflationary cosmology}-\cite{Status of non-Riemannian cosmology}.

The organization of the paper is as follows: In section \ref{The model}, following the proposed model in \cite{Langvik} and \cite{Higgs inflation in Einstein-Cartan gravity} we motivate and explain the inclusion of a scalar sector together with a slight generalization of the gravitational model (\ref{GravitationalAction}) characterized by the addition of three scalar functions, moreover, we show how the torsion can be fully written in terms of the 
proposed scalar. We end up the section with a discussion about the projective and conformal transformations which will be our main contribution to this section, here we will introduce what we call \enquote{extended conformal transformation} and explain how this relates to the transformation originally used in \cite{Langvik}, we then will discuss the usefulness of this new approach. In section \ref{Einstein Frame} we show how, by using the proposed extended conformal transformation we can recover the same Einstein frame action as in \cite{Langvik} and explain how this two methods differ from one another. In section \ref{Jordan Frame} we focus on the unexplored Jordan frame of the model showing how this frame leads to an action which corresponds to a Brans-Dicke action with $\omega_{BD}\neq 1$, this Brans-Dicke model in the Jordan frame will be the core of our contribution and will serve as a way to compare the results found in \cite{Langvik} and \cite{Higgs inflation in Einstein-Cartan gravity} with those found in this paper. In section 
\ref{Cosmological background} we show the cosmological dynamics of the Jordan frame of the model by choosing the traditional FLRW metric. We study the dynamics of the model in both our canonical and our non canonical actions just to show their equivalence, we also study the stability of such models in order to explore whether or not the inflationary fixed points exist. This section purpose is to verify that the Jordan frame of the action exhibits the same dynamical stability as the one in the Einstein frame. Section \ref{Scale invariant case}
contains the scale invariant case of our generic model. We motivate this choice and show that it can indeed produce a good description of the inflationary phase of the universe by being in agreement with the Planck measurements \cite{Planck 2018 results. X. Constraints on inflation}. Finally section \ref{Conclusions} contains our conclusions.

\section{The model}\label{The model}
\subsection{Scalar sector.} Scalar fields are widely and popularly employed in cosmology to explain some phenomena like dark matter \cite{Cold and fuzzy dark matter}-\cite{Strong Constraints on Fuzzy Dark Matter from Ultrafaint Dwarf Galaxy Eridanus}, dark energy \cite{Dynamics of dark energy}-\cite{An Alternative to quintessence} and inflation \cite{Cosmological Consequences of a Rolling Homogeneous Scalar Field}-\cite{Hybrid inflation} nonetheless, note that almost all power-law models of inflation $(V\propto \phi^{n})$ have been already ruled out by Planck results \cite{Planck 2018 results. X. Constraints on inflation}. In fact the first and dynamically easiest model of inflation includes a minimally coupled scalar field in the so called slow roll approximation. However, before the discovery of a fundamental scalar, the Higgs boson \cite{Higgs}, there was some controversy on whether or not the scalars used in cosmology really existed. Even with the discovery of the Higgs boson, the inflaton field is still unknown but the fact that a fundamental scalar has been discovered is a positive point in the discussion. 

In fact, the Higgs boson  has been considered as the scalar field responsible for inflation in the so called \enquote{Higgs inflation} \cite{The Standard Model Higgs boson as the inflaton} by identifying the scalar field in the Brans-Dicke model \cite{Jordan-Brans-Dicke Theory} as the Higgs boson. The original version of Higgs inflation gives the same predictions for the tensor-to-scalar ratio and scalar spectral index as the Starobinsky model \cite{Starobinsky} which is in great agreement with experiments.

In the present work, we will consider the model proposed in \cite{Langvik} and \cite{Higgs inflation in Einstein-Cartan gravity}. A general, non-minimally coupled, Brans-Dicke like model whose action is given by 
\begin{equation}\label{GenericAction}
    S=\int d^{4}x\sqrt{-g}\left[\frac{1}{2}F(h)R+\frac{1}{2}G(h)\mathcal{R}+Y(h)\partial_{\mu}\epsilon^{\mu\nu\rho\sigma}T_{\nu\rho\sigma}-\frac{1}{2}\partial_{\mu}h\partial^{\mu}h-V(h) \right].
\end{equation}
Where, $F(h)$, $G(h)$ and $Y(h)$ are generic functions of the scalar field $h$.

We will now proceed as follows, with the definitions (\ref{RicciTensor})-(\ref{TorsionScalar}) we can write the action (\ref{GenericAction}) fully in terms of torsion and the Levi-Civita Ricci scalar $\mathring{R}$. Once the action has the mentioned form, we can find out the equations of motion for the torsion whose solution will then be re-plugged into the original action (\ref{GenericAction}) obtaining an effective metric theory. This procedure has been previously carried out in several works by different authors like those in \cite{Higgs inflation in Einstein-Cartan gravity}, \cite{BRANS-DICKE TYPE MODELS WITH TORSION} and \cite{Higgs-Dilaton inflation in Einstein-Cartan gravity}. In this part of the paper we will closely follow the analysis done in \cite{Langvik} for us to explicitly show the differences and similarities between the projective transformation and the extended conformal transformation originally introduced in \cite{ON THE RENORMALIZATION OF MODELS OF QUANTUM FIELD THEORY IN AN EXTERNAL GRAVITATIONAL FIELD WITH TORSION} and \cite{BRANS-DICKE TYPE MODELS WITH TORSION}. 

\subsubsection{Solving for torsion.} Taking the variation of the action (\ref{GenericAction}) with respect to torsion, we find the following equations of motion 
\begin{equation}\label{Equationofmotionoftorsion}
\begin{split}
&2Fg_{\alpha[\beta}T_{\gamma]}+FT_{\alpha\beta\gamma}+G\epsilon_{\alpha\beta\gamma}^{\ \ \ \ \mu}T_{\mu}+...\\
    &...+\frac{1}{2}G\epsilon_{\alpha\beta}^{\ \ \ \mu\nu}T_{\gamma\mu\nu}+2g_{\alpha[\beta}\partial_{\gamma]}F+\epsilon_{\alpha\beta\gamma}^{\ \ \ \ \mu}\partial_{\mu}(G-Y)=0.
\end{split}
\end{equation}
We can now find the solution of these equations of motion by contracting with the metric tensor finding that 
\begin{equation}\label{torsionsolution2}
    T_{\alpha\beta\gamma}=t_{1} g_{\alpha[\beta}\partial_{\gamma]}h+t_{2}\epsilon_{\alpha\beta\gamma}^{\ \ \ \ \mu}\partial_{\mu}h,
\end{equation}
together with the two contractions 
\begin{equation}\label{torsionsolution1}
    T_{\mu}=-\frac{3}{2}t_{1}\partial_{\mu}h \quad \mbox{and} \quad \hat{T}_{\mu}=t_{2}\partial_{\mu}h.
\end{equation}
Where 
\begin{equation}\label{t1t2}
    t_{1}=\frac{FF'+G(G'-Y')}{F^{2}+G^{2}}, \quad 
    t_{2}=\frac{GF'-F(G'-Y')}{F^{2}+G^{2}} \quad \mbox{and}\quad F'=\frac{\partial F}{\partial h}.
\end{equation} 
We can furthermore relate these two equations as 
\begin{equation}
    t_{1}=\left(\frac{FF'+G(G'-Y')}{GF'-F(G'-Y')}\right)t_{2}.
\end{equation}
\subsection{Projective transformation vs conformal transformation} 

\subsubsection{Projective transformation.}Whenever one has a theory of gravity in which one considers the metric $g_{\mu\nu}$ and the connection $\Gamma^{\lambda}_{\ \mu\nu}$ to be independent (Palatini formulation), there exists a transformation on the connection which leaves the Ricci scalar invariant. This transformation is called the projective transformation which was originally studied by Einstein himself \cite{The Meaning of Relativity}. This transformation has the following form: 
\begin{equation}\label{Projectivetransformation}
    \Gamma^{\lambda}_{\ \mu\nu}\xrightarrow[]{}\Gamma^{\lambda}_{\ \mu\nu}+\delta^{\lambda}_{\nu}A_{\mu}.
\end{equation}
Where $A_{\mu}$ is an arbitrary vector. Under such transformation, the Riemann and the Ricci tensors transform as 
\begin{equation}
    \mathring{R}_{\mu\nu\lambda}^{\ \ \ \ \rho}\xrightarrow[]{}\mathring{R}_{\mu\nu\lambda}^{\ \ \ \ \rho}+2\partial_{[\mu}A_{\nu]}\delta^{\lambda}_{\rho} \quad \mbox{and}\quad \mathring{R}_{\mu\nu}\xrightarrow[]{}\mathring{R}_{\mu\nu}+2\partial_{[\mu}A_{\nu]}.
\end{equation} 
From here it is clear that the Ricci scalar is invariant under such transformation. Indeed one can extend this transformation when considering a theory of gravity containing also torsion. In this case the Riemann tensor transforms as in \cite{Langvik}
 \begin{equation}
    R_{\mu\nu\lambda\rho}\xrightarrow[]{}R_{\mu\nu\lambda\rho}+g_{\mu\nu}\left(2\nabla_{[\lambda}A_{\rho]}+T^{\alpha}_{\ \lambda\rho}A_{\alpha}\right).
\end{equation} 
It is therefore easy to see that this new form of the Riemann tensor leads to an invariant Ricci scalar. Note also that, under the projective transformation, the Nieh-Yan invariant and the Holst term are also invariant.

\subsubsection{(Extended) Conformal transformation.} Usually, in the context of gravitational theories, cosmology, and specially inflation, the conformal transformation is a rescaling of the metric that is used to go from the so called \enquote{Jordan frame} to the \enquote{Einstein frame}. The former is in general described by an action which is not linear in the Ricci scalar and the latter is described by an action which is indeed linear in the Ricci scalar and thus, in principle, the Einstein frame is generally and easier frame to work on.

The conformal transformation on the metric is given by
\begin{equation}\label{Conformaltransformation}
    \Tilde{g}_{\mu\nu}=e^{2\Omega(x)}g_{\mu\nu}.
\end{equation}
Where $\Omega(x)$ is the conformal factor.
Under the transformation (\ref{Conformaltransformation}) the Levi-Civita Ricci scalar transforms as 
\begin{equation}\label{Ricciscalarconformaltransformation}
    \mathring{R}=e^{-2\Omega}(\Tilde{\mathring{R}}+6\Tilde{\Box}\Omega-6\Tilde{g}^{\mu\nu}\partial_{\mu}\Omega\partial_{\nu}\Omega).
\end{equation}

However, now that our Ricci scalar is not only composed of its Levi-Civita part but also contains a torsion contribution, it is not obvious how the conformal transformation should apply on the Ricci scalar. In other words, the question is now, how does the torsion transforms under conformal transformations?
Sadly, there is no universal answer, rather the transformation of the torsion is, in principle, a choice one makes.  
The choice we made in this work is that the torsion transforms in a way such that the full Riemann tensor (not to be confused with the Levi-Civita Riemann tensor) is invariant under the conformal transformation. In other words, that the Ricci scalar transforms as 
\begin{equation}
    \Tilde{R}=e^{-2\Omega}R.   
\end{equation}
After some reverse engineering, it is easy to confirm that, if the torsion transforms as 
\begin{equation}\label{torsiontransformation}
    \Tilde{T}^{\lambda}_{\ \mu\nu}= T^{\lambda}_{\ \mu\nu}-\delta^{\lambda}_{\mu}\Omega_{,\nu}+\delta^{\lambda}_{\nu}\Omega_{,\mu}.
\end{equation}
we achieve the desired transformation of the Ricci scalar. We can also write this transformation in a more general way. In addition to (\ref{Conformaltransformation}) we also have a transformation on the connection 
\begin{equation}\label{Connectiontransformation}
    \Tilde{\Gamma}^{\lambda}_{\,\, \mu\nu}=\Gamma^{\lambda}_{\,\, \mu\nu}+\delta^{\lambda}_{\nu}\Omega_{,\mu}.
\end{equation}
Both (\ref{Conformaltransformation}) and (\ref{torsiontransformation}) compose what we call \enquote{extended conformal transformation} which has been previously studied \cite{ON THE RENORMALIZATION OF MODELS OF QUANTUM FIELD THEORY IN AN EXTERNAL GRAVITATIONAL FIELD WITH TORSION} and \cite{BRANS-DICKE TYPE MODELS WITH TORSION}.

In principle it would seem that the projective transformation for the connection (\ref{Projectivetransformation}) is equivalent to the extended conformal transformation, indeed the extended conformal transformation (\ref{Connectiontransformation}) is a special case of (\ref{Projectivetransformation}) in which $A_{\alpha}=\partial_{\alpha}\Omega$ with $\Omega$ being a scalar field and with the added ingredient that the metric also transforms, as in (\ref{Conformaltransformation}). This case makes the Riemann tensor invariant and thus the seemingly equivalence between the two aforementioned transformations. 
The key difference between these two transformations relies on the fact that the projective transformation is an invariance the system possesses on its own. On the other hand, the extended conformal transformation is a consequence of the fact that the torsion must also transform under the conformal transformation. Moreover, in the context of the projective transformation, nor the metric neither the torsion transform. The connection is the fundamental object that transforms. In contrast, when applying the extended conformal transformation, both the metric and the torsion transform hence the difference between these two transformations. Finally, note also that both the Nieh-Yan invariant and the Holst term are invariant under both the projective transformation and under the extended conformal transformation. 

Some advantages of using the extended conformal transformation is that, in principle, one can choose how the torsion transforms. Depending on what one is working on, there could be several different ways in which the torsion may transform. On the other hand, when using the projective transformation, the transformation is already set and cannot be, in general, changed. The extended conformal transformation offers a wider freedom than the projective transformation. 
\section{Einstein Frame}\label{Einstein Frame}

This section will serve as a comparison of methods. On \cite{Langvik} the authors performed a conformal transformation at the level of the full Ricci scalar $R$, in principle this should not be correct because the full Ricci scalar contains torsion contributions and hence, taking the conformal transformation at this level (without splitting the Ricci scalar into its Levi-Civita and torsion components) implicitly assumes that the torsion terms inside $R$ transform in the same way as the Levi-Civita Ricci scalar $\mathring{R}$. In order to solve this \enquote{problem} the authors rely on the fact that the full Ricci scalar is invariant under the projective transformation and hence, in principle, they didn't need to separate the torsion and Levi-Civita components of the Ricci scalar. On the other hand, in our method, we perform the conformal transformation at the level of the Levi-Civita Ricci scalar. At this level, the curvature and torsion contributions to $R$ are already splitted and then, the question on how the torsion transforms needs to be answered. At this point, we propose the extended conformal transformation and thus we set a transformation for the torsion and the metric. As mentioned before, we made the choice on how the torsion needs to transform in order to recover the Einstein frame in \cite{Langvik} but, maybe some authors, in some other work and context, choose a different transformation that leads to a different action.

Under the extended conformal transformation with $\Omega=F$, the original action (\ref{GenericAction})  takes the form 
\begin{equation}\label{Einsteinframeaction}
    S=\int d^{4}x\sqrt{-g}\left[\frac{1}{2}R+\frac{1}{2}\frac{G(h)}{F(h)}\mathcal{R}-\frac{3}{2}\hat{T}^{\mu}\frac{\partial_{\mu}Y(h)}{F(h)}-\frac{1}{2F(h)}\partial_{\mu}h\partial^{\mu}h-U(h) \right].
\end{equation}
Where the potential is given by $U=\frac{V}{F^{2}}$. This action, as already anticipated is exactly the same found in \cite{Langvik} and thus all the analysis on that work holds. For the sake of completeness, we will only show the final form of the action for us to be able to compare with later on. 

Opening up the Ricci scalar and the Holst term and substituting the torsion in terms of the scalar field using  (\ref{torsionsolution2})-(\ref{torsionsolution1}) we end up with the effective metric action in the Einstein frame
\begin{equation}
\begin{split}
    S=\int d^{4}x\sqrt{-g}\left[\frac{1}{2}\mathring{R}-\frac{1}{2}\left(\frac{1}{F}+\frac{3}{2}\frac{[GF'-F(G'-Y')]^{2}}{F^{2}(F^{2}+G^{2})}\right)\left(\partial_{\mu}h\right)^{2}-U\right].
\end{split}    
\end{equation}
We can therefore write the action in terms of the non-canonical kinetic term contribution $\omega_{E}$ as
\begin{equation}
    S=\int d^{4}x\sqrt{-g}\left[\frac{1}{2}\mathring{R}-\frac{1}{2}\omega_{E}\left(\partial_{\mu}h\right)^{2}-U\right].
\end{equation}
Where 
\begin{equation}
    \omega_{E}=\frac{1}{F}+\frac{3}{2}\frac{[GF'-F(G'-Y')]^{2}}{F^{2}(F^{2}+G^{2})}.
\end{equation}
The final form of the action in the Einstein frame is a traditional Einstein Hilbert term plus the action for a scalar field minimally coupled to gravity but with a non-canonical kinetic term. 
\section{Jordan Frame}\label{Jordan Frame}

In this section we will analyze the less explored Jordan frame of the action and start to compare some of the results found in this frame with the ones found in the Einstein frame in \cite{Langvik}. In the Jordan frame, the action reads 
\begin{equation}\label{Jordanframeaction}
    S=\int d^{4}x\sqrt{-g}\left[\frac{1}{2}F(h)R+\frac{1}{2}G(h)\mathcal{R}-\frac{3}{2}\hat{T}^{\mu}\partial_{\mu}Y(h)-\frac{1}{2}\partial_{\mu}h\partial^{\mu}h-V(h) \right].
\end{equation}
We will now follow the same steps as in the Einstein frame section, we will use the solution of the equations of motion of the torsion in order to obtain an effective metric action. Opening up the Ricci scalar, the Holst term and the torsion scalar using (\ref{torsionsolution2})-(\ref{torsionsolution1}) we have
\begin{equation}\label{JordanScalaraction}
    S=\int d^{4}x\sqrt{g}
   \Bigg[\frac{1}{2}F\mathring{R}+\frac{1}{2}\Bigg[-6Ft_{1}^{2}+\frac{3F}{2}t_{2}^{2}+3F't_{1}+3t_{2}G'-6Gt_{1}t_{2}-3t_{2}Y'-1\Bigg](\partial_{\mu}h)^{2}-V(h)\Bigg].
\end{equation}
Note that, in this calculation we cannot throw away the divergence term in (\ref{RicciTensor}) because of the non-minimal coupling with the scalar field.

Once again, note that after writing the torsion fully in terms of the scalar field, we have obtained an effective metric theory (\ref{JordanScalaraction}) that now contains a non-minimal coupling between the Levi-Civita Ricci scalar and the scalar field together with a non-canonical kinetic term parameterized by $\omega_{J}$ defined as 
\begin{equation}\label{OmegaJ}
    \omega_{J}=6Ft_{1}^{2}-\frac{3F}{2}t_{2}^{2}-3F't_{1}-3t_{2}G'+6Gt_{1}t_{2}+3t_{2}Y'+1.
\end{equation}
Here we can make the first comparison between the actions in the Jordan and in the Einstein frames. In the latter, an Einstein Hilbert term was present and in the former we have an explicit coupling between the scalar function $F$ and the Levi-Civita Ricci scalar. This is exactly what we would expect, by performing the conformal transformation we can get rid of the non-minimal coupling and hence go from one frame to the other. The second comparison between the two frames can be done by comparing the non-canonical kinetic terms contributions $\omega_{E}$ and $\omega_{J}$. Indeed, the transformation from one frame to another can be accounted by the following identifications also found in \cite{Langvik} 

$$F\xrightarrow[]{}1,\quad G\xrightarrow{}\frac{G}{F}, \quad Y'\xrightarrow{}\frac{Y'}{F},$$
with which it is easy to see that $\omega_{J}\xrightarrow[]{}\omega_{E}$. We can then write the Jordan frame action in the following form: 
\begin{equation}\label{NonkanonicalBransDickelikeaction}
   S=\int d^{4}x\sqrt{-g}\Bigg[\frac{1}{2}F\mathring{R}-\frac{1}{2}\omega_{J}\partial_{\mu}h\partial^{\mu}h-V(h)\Bigg].
\end{equation}
Whenever we have $G=Y=0$ we should recover the well known Higgs inflation model \cite{The Standard Model Higgs boson as the inflaton}, indeed whenever we have these two non-minimal couplings equal to zero, both $t_{1}$ and $t_{2}$ vanish and therefore we obtain $\omega_{J}=1$ which gives the correct sign of the kinetic term in the traditional Higgs inflation model. Note also that we can further redefine the scalar field in order to have a canonical kinetic term in terms of a new field $\chi$ as 
\begin{equation}\label{scalarfieldredefinition}
    \frac{\partial \chi}{\partial h}=\sqrt{\omega_{J}}.
\end{equation}
Thus, obtaining the canonical action 
\begin{equation}\label{kanonicalBransDickelikeaction}
    S=\int d^{4}x\sqrt{-g}\Bigg[\frac{1}{2}F\mathring{R}-\frac{1}{2}\partial_{\mu}\chi\partial^{\mu}\chi-V(\chi)\Bigg].
\end{equation}
 One can either choose to work with the \enquote{canonical action} (\ref{kanonicalBransDickelikeaction}) or with the \enquote{non-canonical action} (\ref{NonkanonicalBransDickelikeaction}) and the results shouldn't vary, however, there are certainly big differences. By working on the latter, we need to take into account the explicit appearance of $\omega_{J}$ in the equations of motion and hence, it is not obvious how this explicit non-canonical term affects the equations of motion and their stability \footnote{Recall that, in general, $\omega_{J}$ depends on $h$. }. By working on the former, one ensures to have the Klein-Gordon equation for the scalar field $\chi$ but the analysis on the original scalar field $h$ turns out to be not so transparent. As both scalar fields are related by $\omega_{J}$, the analysis of the action (\ref{kanonicalBransDickelikeaction}) is extremely model dependent, it losses its generic properties very quickly. This argument may sound strange because in the non-canonical action (\ref{NonkanonicalBransDickelikeaction}) there appears to be the same problem as $\omega_{J}$ appears explicitly but in the next section we will see how there is a generic fixed point that is model independent.
\section{Cosmological background and dynamical analysis of the model}\label{Cosmological background}

In this section we are interested in the study of the stability of the model to ensure it possess the necessary fixed points in order to model the early expansion of the universe, this is once more traditionally done in the Einstein frame of the theories in which gravity is just given by the Einstein Hilbert term, however, in our present case it is not obvious how the non-canonical kinetic term together with the non-minimal coupling between the scalar field and the curvature may affect the stability of the system. The reader may be interested in taking a look at \cite{Rinaldi}, here the authors use dynamical analysis tools in both the Einstein and Jordan frames and show how, in principle, the two frames look very different from one another. In our present case this difference is not so notable because all the dynamics are encoded in the scalar field but we think it is worth to explore the dynamics in the Jordan frame. 

The choice of background we will work on is the FLRW metric whose line element is given by $ds^{2}=-dt^{2}+a^{2}(t)dx_{i}dx^{i}$, where $a(t)$ is the scale factor. 
As a final consideration before moving on to the calculations, we need to recall that in the FLRW background the Levi-Civita Ricci scalar is given by 
\begin{equation}\label{FRLWRicciscalar}
    \mathring{R}=12H^{2}+6\dot{H}.
\end{equation}
Where $H$ is the Hubble factor defined as $H=\dot{a}/a$. Notice that (\ref{GenericAction}) does not contain second derivatives of the Hubble parameter and hence, as previously stated, the dynamics of the model are encoded fully on the scalar field \footnote{It is clear that the scale factor is also dynamical but we are interested in the Hubble parameter, not the scale factor.}. 

\subsection{Canonical action dynamical analysis}

The equations of motion corresponding to  (\ref{kanonicalBransDickelikeaction}) are 
\begin{equation}\label{equationsofmotioncanonicalaction}
    3FH^{2}=\frac{1}{2}\dot{\chi}^{2}+V(\chi)-3H\dot{F},
\end{equation}
\begin{equation*}
    2F\dot{H}=-\dot{\chi}^{2}-\ddot{F}+H\dot{F},
\end{equation*}
\begin{equation*}
    \ddot{\chi}+3H\dot{\chi}+V_{,\chi}-3(\dot{H}+2H^{2})F_{,\chi}=0.
\end{equation*}
Where $,_{\chi}$ represents the derivative with respect to $\chi$.
Lets now treat this system of equations as a dynamical system for which we will find its fixed points and then study the stability around them.
By setting all the time derivatives to zero in (\ref{equationsofmotioncanonicalaction}) we find that the fixed points of the system are given by the solutions to the following equation 
\begin{equation}\label{FixedPointEquation}
    3FH^{2}=V(\chi).
\end{equation}

\subsubsection{Unstable fixed point.} The first fixed point is given by
\begin{equation}\label{Unstablefixedpoint}
    (H,\chi,\dot{H},\dot{\chi})=(H,0,0,0).
\end{equation}
The fact that $\chi=0$ is a solution of (\ref{FixedPointEquation}) is based on the assumption of only considering polynomial like potentials\footnote{Generic potentials may not solve (\ref{FixedPointEquation}) at $\chi=0$.}. 
As we would like to have some inflationary observables, we change variables from time to number of $e$-folds as follows 
\begin{equation}
    H=\frac{\dot{a}}{a}=\frac{d}{dt}\ln a=\frac{dN}{dt} \quad \mbox{where we defined} \quad N=\ln a.
\end{equation}
Consequently, the scalar equation of motion is given by 
\begin{equation}
    \chi_{(N)}H_{(N)}H+H^{2}\chi_{(N\ N)}+3H^{2}\chi_{(N)}+V_{,\chi}-3(H_{(N)}H+2H^{2})F_{,\chi}=0,
\end{equation}
where $_{(N)}$ represents derivative with respect to the $e$-folding number and $_{(N\ N)}$ represent the second derivative.
In a more convenient form we have
\begin{equation}
    \chi_{(N \ N)}+3\chi_{(N)}+\frac{\chi_{(N)}H_{(N)}}{H}+\frac{V_{,\chi}}{H^{2}}-3\frac{H_{(N)}F_{,\chi}}{H}-6F_{,\chi}=0.
\end{equation}
Notice that the fixed point equation (\ref{FixedPointEquation}) in terms of the $e$-folding number is given by
\begin{equation}
     H^{2}=\frac{V_{,\chi}}{6F_{,\chi}}.
\end{equation}
Let us first focus on the fixed point represented by an arbitrary value of the Hubble parameter and a null scalar field. The equation of motion around the aforementioned fixed point (\ref{Unstablefixedpoint}) reads 
\begin{equation}\label{Equationofmotionunstablefixedpoint}
    \chi_{(N \ N)}+3\chi_{(N)}-6F_{,\chi}=0.
\end{equation}
As we have already mentioned, the exact behavior around the unstable fixed point (the solution to (\ref{Equationofmotionunstablefixedpoint})) is extremely model dependent. However, just by looking at this differential equation we can immediately notice that the solution of such equation should be a combination of increasing and decreasing modes and therefore we can argue that the fixed point  (\ref{Unstablefixedpoint}) is a saddle point, an unstable fixed point. 
Indeed in the limit of vanishing $G$ and $Y$ where we have $\omega_{J}=1$ and taking $F\propto h^{2}$ the behavior of the scalar field is given by
\begin{equation}
    h_{(N \ N)}+3h_{(N)}-12h=0,
\end{equation}
whose solution is indeed a combination of decreasing and increasing modes and thus a saddle point.

\subsubsection{Stable fixed point.}
The second fixed point of the system is given by
\begin{equation}\label{stablefixedpoint}
    (H,\chi,H_{(N)},\chi_{(N)})=\left(\sqrt{\frac{V_{,\chi}}{6F_{,\chi}}},\chi,0,0\right).
\end{equation}
Around this fixed point, the behavior of the scalar field is given by 
\begin{equation}
    \chi_{(N N)}+3\chi_{(N)}=0 \quad \xrightarrow[]{}\quad \chi=C_{1}e^{-3N}+C_{2}.
\end{equation}
Which is an attractor solution and therefore a stable fixed point. The scalar field tends to a constant. Notice also that this second fixed point is not so much model dependent, it is clear that the above differential equation has a decreasing exponential behavior and that this solution is not dependent on the details of the model. 

\subsection{Non-canonical action dynamical analysis}
We will now consider the non-canonical action (\ref{NonkanonicalBransDickelikeaction}) to explicitly show that both the canonical and the non-canonical actions are dynamically equivalent. 
The scalar equation of motion corresponding to (\ref{NonkanonicalBransDickelikeaction}) is 
\begin{equation}\label{Scalarequationofmotion}
    6F'H^{2}+3F'\dot{H}-\frac{1}{2}\omega_{J}'\dot{h}^{2}-V'-\omega_{J}(\ddot{h}+3H\dot{h})=0.
\end{equation}
Once again, in order to find the fixed points of this dynamical equation we set the time derivatives to zero obtaining
\begin{equation}
    6F'H^{2}-V'=0 \xrightarrow{}H^{2}=\frac{V'}{6F'}.
\end{equation}
In terms of the $e$-folding number, the equation of motion for the scalar field takes the following form 
\begin{equation}\label{Scalarequationofmotionefolding}
    6F'H^{2}+3F'HH_{(N)}-\frac{1}{2}\omega_{J}'h^{2}_{(N)}H^{2}-V'-\omega_{J}(HH_{(N)}h_{(N)}+H^{2}h_{(N N)}+3H^{2}h_{(N)})=0.
\end{equation}
\subsubsection{Unstable fixed point.}Arranging the terms in a more convenient way we have 
\begin{equation}
    -h_{(N\ N)}\omega_{J}-3\omega_{J} h_{(N)}+6F'+3F'\frac{H_{(N)}}{H}-\frac{1}{2}\omega_{J}'h^{2}_{(N)}-\frac{V'}{H^{2}}-\omega_{J} h_{(N)}\frac{H_{(N)}}{H}=0.
\end{equation}
Around the first fixed point $(H,h,H_{(N)},h_{(N)})=(H,0,0,0)$ the equation of motion takes the form 
\begin{equation}
    \omega_{J} h_{(N\ N)}+ 3\omega_{J} h_{(N)}-6F'=0.
\end{equation}
Here, once again the exact behavior of the scalar field is model dependent but, as before, we can argue that this fixed point is unstable. For the case $G=Y=0$ we have already seen that $\omega_{J}=1$ and taking $F\propto h^{2}$ the equation of motion around the fixed point reads 
\begin{equation}
    h_{(N\ N)}+3h_{(N)}-12h=0,
\end{equation}
whose solution is indeed a saddle fixed point. 

\subsubsection{Stable fixed point.} The second fixed point is given by $(H,h,H_{(N)},h_{(N)})=\left(\sqrt{\frac{V'}{6F'}},h,0,0\right)$ which then leads to an equation of motion given by 
\begin{equation}
    \omega_{J} h_{(N N)}+3\omega_{J} h_{(N)}=0,
\end{equation}
whose solution is insensitive to the value of $\omega_{J}$. The solution to this equation is given by 
\begin{equation}
    h=c_{1}e^{-3N}+c_{2}.
\end{equation}
This solution tends to a constant and therefore to an attractor fixed point.

We have now shown that our effective metric theory (\ref{JordanScalaraction}) whether in its canonical or non-canonical form contains the two fixed points needed for inflation to possibly take place. One unstable fixed point in which the universe will inflate and a stable fixed point where the system will tend to go to. This stable fixed point is in the region of graceful exit from inflation.
Moreover, we have found that the existence of a stable fixed point is very incentive to the details of the model. In hindsight, this was expected as the equation of motion of a minimally coupled scalar field in a FLRW background has a stable fixed point around its minimum\footnote{This minimum is achieved through the Hubble friction term.}, however it was not obvious that this behavior will be preserved in our effective metric theory. In addition, the fact that the stable fixed point exists means that our effect metric theory (\ref{JordanScalaraction}) has a graceful exit from the inflationary epoch that needs to then give rise to GR\footnote{After inflation and the possible reheating of the universe we need to ensure that we recover a non-accelerating cosmology and that we have the GR theory in order to be in agreement with the $\Lambda$CDM model.}. On the contrary, the unstable fixed point is much more model dependent and hence a generic analysis begins to be not 
so useful. For this we will study an specific model in the next section. Moreover we have now fully demonstrated that the effective metric action, whether in its canonical or non-canonical form has the correct stability conditions in order to possibly describe inflation. As already mentioned, this analysis tends to be done in the Einstein frame but here we have found that up until this point, the results in the Jordan frame are equivalent to those in the Einstein one.  

\section{Scale invariant case}\label{Scale invariant case}

There are certainly many ways to write down a scale invariant theory of gravity, some of them can be found in \cite{Racioppi} and \cite{Scale invariance dynamically induced Planck scale and inflation in the Palatini formulation} but in this section we will stick to the simplest case which is given by setting all our scalar functions $F$, $G$ and $Y$ to be proportional to $h^{2}$. Furthermore, we will consider a quartic potential of the form $V=\frac{1}{4}\lambda h^{4}$, where $\lambda$ is a coupling constant, hence the action.\footnote{Recall that the scalar field has units of energy.}
\begin{equation}\label{Scaleinvariantanswer}
    S=\int d^{4}x\sqrt{-g}\left[\frac{1}{2}fh^{2}R+\frac{1}{2}gh^{2}\mathcal{R}+yh^{2}\partial_{\mu}\epsilon^{\mu\nu\rho\sigma}T_{\nu\rho\sigma}-\frac{1}{2}\partial_{\mu}h\partial^{\mu}h-\frac{1}{4}\lambda h^{4} \right].
\end{equation}
Here, $f$, $g$ and $y$ are all dimensionless couplings. In principle we would need to repeat the process of solving for torsion but thanks to the generic analysis previously carried out in \cite{Langvik} the action (\ref{Scaleinvariantanswer}) can be rewritten as 
\begin{equation}\label{BDScaleinvariantanswer}
    S=\int d^{4}x\sqrt{-g}\left[\frac{1}{2}fh^{2}\mathring{R}-\frac{1}{2}\omega_{SI}\partial_{\mu}h\partial^{\mu}h-\frac{1}{4}\lambda h^{4} \right].
\end{equation}
Where
\begin{equation}\label{OmegaJscaleinvariant}
\begin{split}
    \omega_{SI}
    =&+6f\left(\frac{2f^{2}+2g^{2}-2yg}{f^{2}+g^{2}}\right)^{2}-\frac{3f}{2}\left(\frac{2fy}{f^{2}+g^{2}}\right)^{2}\\
    &-6f\left(\frac{2f^{2}+2g^{2}-2yg}{f^{2}+g^{2}}\right)-6g\left(\frac{2fy}{f^{2}+g^{2}}\right)\\
    &+6g\left(\frac{2f^{2}+2g^{2}-2yg}{f^{2}+g^{2}}\right)\left(\frac{2fy}{f^{2}+g^{2}}\right)\\
    &+6y\left(\frac{2fy}{f^{2}+g^{2}}\right)+1.
\end{split}
\end{equation}
and
\begin{equation}\label{t1t2scaleinvariant}
    t_{1}=\frac{1}{h}\left(\frac{2f^{2}+2g^{2}-2yg}{f^{2}+g^{2}}\right), \quad 
    t_{2}=\frac{1}{h}\left(\frac{2fy}{f^{2}+g^{2}}\right).
\end{equation} 
It is clear that $\omega_{SI}$ is a very complicated algebraic expression, however, notice that it only depends on the three couplings $f$, $g$ and $y$ hence this non-canonical function is just a constant, therefore the relation between the two scalar fields (\ref{scalarfieldredefinition}) is very simple.  
Indeed by taking $\chi=\sqrt{\omega_{J}}h$ we can go from the non-canonical representation (\ref{kanonicalBransDickelikeaction}) to the canonical one (\ref{NonkanonicalBransDickelikeaction}).
\subsection{Slow roll}

In this section we will investigative the inflationary phase of our scale invariant model (\ref{Scaleinvariantanswer}). In order to do this, we first try to set some bounds on the free parameters of our model. 
Starting from $(\ref{Scalarequationofmotionefolding})$ with $\omega_{J}=\omega_{SI}$

\begin{equation}
    6F'H^{2}+3F'HH_{(N)}-\frac{1}{2}\omega_{SI}'h^{2}_{(N)}H^{2}-V'-\omega_{SI}(HH_{(N)}h_{(N)}+H^{2}h_{(N N)}+3H^{2}h_{(N)})=0,
\end{equation}
dividing by $H^{2}$
and taking the common factor $H_{(N)}/H$ out 

\begin{equation}
    \frac{H_{(N)}}{H}\left(3F'-\omega_{SI}h_{(N)}\right)+6F'-\frac{1}{2}\omega_{SI}'h^{2}_{(N)}-\frac{V'}{H^{2}}-\omega_{J}\left(h_{(N N)}+3h_{(N)}\right)=0,
\end{equation}
and defining the slow roll parameter $\epsilon$ as
\begin{equation}
    \epsilon=-\frac{\dot{H}}{H^{2}}=-\frac{H_{(N)}}{H},
\end{equation} 
we have the following expression:

\begin{equation}
    \epsilon=-\frac{H_{(N)}}{H}=\frac{h_{(N N)}\omega_{SI}+3\omega_{SI} h_{(N)}-6F'+\frac{1}{2}\omega_{SI}'h_{(N)}^{2}+\frac{V'}{H^{2}}}{\omega_{SI} h_{(N)}-3F'}.
\end{equation}
For our particular choice of $F$ and $V$ we then have 
\begin{equation}
    \epsilon=-\frac{H_{(N)}}{H}=\frac{h_{(N N)}\omega_{SI}+3\omega_{SI} h_{(N)}-12fh+\lambda h^{3}/4H^{2}}{\omega_{SI} h_{(N)}-6fh}.
\end{equation}
Setting the end of inflation at the time when $\epsilon\approx 1$, we can then calculate the number of $e$-folds up to the end of inflation.
Recalling that the fixed point that ends inflation is the stable fixed point, we can then expand the scalar field and the Hubble parameter around the unstable fixed point as 
\begin{equation}
    h\xrightarrow[]{}h_{0}e^{\Delta N} \quad \mbox{and} \quad H\xrightarrow[]{}H_{0}. \quad \mbox{where} \quad \Delta N=N_{e}-N_{0}.
\end{equation}
The subscript subzero refers to the start of inflation while $N_{e}$ refers to the time when $\epsilon=1$.
Consequently, at the end of inflation we have
\begin{equation}
    1=\frac{h_{0}e^{\Delta N}\omega_{SI}+3\omega_{SI} h_{0}e^{\Delta N}-12fh_{0}e^{\Delta N}+\lambda h_{0}^{3}e^{3\Delta N}/4H^{2}}{\omega_{SI} h_{0}e^{\Delta N}-6fh_{0}e^{\Delta N}}.
\end{equation}
Solving for the number of $e$-folds we finally obtain

\begin{equation}\label{efoldings}
    \Delta N=\frac{1}{2}\ln \left[ \frac{4H_{0}^{2}}{\lambda h_{0}^{2}}\left(6f-3\omega_{SI}\right)\right].
\end{equation}
This result resembles very much that one found in \cite{Rinaldi}. Indeed, Rinaldi's et al. result can be obtained from our expression (\ref{efoldings}) whenever 
\begin{equation}\label{conditionequivalencerinaldi}
    22f-12\omega_{SI}+3=0.
\end{equation}
We can further develop this equality between results by taking the limit in which $G=Y=0$. In this limit, $\omega_{SI}$ is given by 
\begin{equation}\label{Omegajrinaldi}
\begin{split}
    \omega_{SI}=12f+1.
\end{split}
\end{equation}
From where it follows that $f\approx \frac{1}{60}$. 
Notice also that we can immediately set some limits  on the couplings by requiring that the argument of the logarithm in (\ref{efoldings}) is positive 
\begin{equation}\label{Logarithmcondition}
    3f-\omega_{SI}<0.
\end{equation}
In order to have a more transparent condition, we will take the limit of vanishing $g$ and $y$, thus the condition (\ref{Logarithmcondition}) becomes 
\begin{itemize}
    \item Case $f\neq 0$, $g=y=0$
\end{itemize}
\begin{equation}\label{bound}
    \omega_{SI}=12f+1\xrightarrow[]{\mbox{positive} \quad \mbox{log} \quad \mbox{condition}}f<\frac{1}{9}
\end{equation}
thus, the bounds set by the positivity of the logarithm are compatible with those found when comparing with Rinaldi's et al. result.  
Notice that the previous calculation that resulted in the expression (\ref{efoldings}) is just a way to set some constraints on the free parameters of the theory (the couplings), this is not in any way the expression for the $e$-folding number that is used in the inflationary predictions. 
\subsection{Observables}

In the final section of this paper we show the inflationary observables coming from the scale invariant case (\ref{Scaleinvariantanswer}) of the generic model (\ref{GenericAction}). We show this procedure in the Jordan frame rather than in the traditional Einstein frame. In fact, working in the Jordan frame is less common and the reader may find an example of the following analysis by looking at \cite{Higgs Inflation with a Gauss-Bonnet term in the Jordan Frame}. At the end of the section we once again compare with results found in the Einstein frame.\\
In the Jordan frame, the  general action (\ref{GenericAction}) reads
\begin{equation}\label{scaleinvariantactionjordanframe}
    S=\int d^{4}x\sqrt{-g}\Bigg[\frac{1}{2}F\mathring{R}-\frac{1}{2}\partial_{\chi}h\partial^{\chi}h-V\Bigg].
\end{equation}
The equations of motion for such model are 
\begin{equation}\label{Friedmanequationsscaleinvariant1}
    3FH^{2}=\frac{1}{2}\dot{\chi}^{2}+V(\chi)-3H\dot{F},
\end{equation}
\begin{equation}\label{Friedmanequationsscaleinvariant2}
    2F\dot{H}=-\dot{\chi}^{2}-\ddot{F}+H\dot{F},
\end{equation}
\begin{equation}\label{Scalarequationsscaleinvariant}
    \ddot{\chi}+3H\dot{\chi}+V_{,\chi}-3(\dot{H}+2H^{2})F_{,\chi}=0.
\end{equation}
From here, it is easy to define the slow roll parameter $\epsilon_{0}$ by using the second Friedmann equation (\ref{Friedmanequationsscaleinvariant2})
\begin{equation}
    \epsilon_{0}=-\frac{\dot{H}}{H^{2}}=\frac{\dot{\chi}^{2}}{2FH^{2}}+\frac{\ddot{F}}{2FH^{2}}-\frac{\dot{F}}{2FH}.
\end{equation}
We further define more slow roll parameters as 
\begin{equation}\label{Furtherslowroll}
    \eta_{0}=\frac{-\dot{F}}{2HF}, \quad \mbox{and} \quad \eta_{1}=\frac{\dot{\eta_{0}}}{H\eta_{0}},
\end{equation}
In general, we can define any further slow roll parameter as 
\begin{equation}
    \epsilon_{n}=\frac{\dot{\epsilon}_{n-1}}{H\epsilon_{n-1}}.
\end{equation}
It is therefore easy to show that both the kinetic term $\dot{\chi}^{2}$ and the potential $V$ can be fully written in terms of the slow roll parameters previously defined 
\begin{equation}\label{kineticterm}
    \dot{\chi}^{2}=H^{2}F\left[2\epsilon_{0}-2\eta_{0}+2\eta_{0}(\eta_{1}-2\eta_{0}-\epsilon_{0})\right],
\end{equation}
\begin{equation}\label{potentialterm}
    V=H^{2}F\left[3-\epsilon_{0}-5\eta_{0}-\eta_{0}\left(\eta_{1}+\eta_{0}+2\epsilon_{0}\right)\right].
\end{equation}
We now proceed to calculate the number of $e$-folds as follows 
\begin{equation}
    N=\int_{t_{0}}^{t} Hdt=\int_{\chi_{0}}^{\chi} \frac{H}{\dot{\chi}}d\chi,
\end{equation}
which, after using the first order approximation of (\ref{kineticterm}) becomes 
\begin{equation}
    N=\int_{\chi_{0}}^{\chi}\frac{1}{\sqrt{2F\epsilon_{0}}}d\chi=\frac{1}{\sqrt{2f\epsilon_{0}}}.
\end{equation}
Note that in the last equation we have specialized our results to our scale invariant case $F=f\chi^{2}$.
Similarly, we can define the slow roll parameter $\eta_{0}$ as follows 
\begin{equation}
    \eta_{0}=\frac{-F_{,\chi}\dot{\chi}}{2HF}=-\frac{\dot{\chi}}{H\chi}=-\frac{1}{N}.
\end{equation}
Finally, we have that the two slow roll parameters we are interested in are related to the $e$-folding number as 
\begin{equation}
    \epsilon_{0}=\frac{1}{2N^{2}} \quad \mbox{and} \quad \eta_{0}=-\frac{1}{N}.
\end{equation}
Here we have assumed that $f\approx 1$. We can now calculate the scalar spectral index $n_{s}$ and the tensor-to-scalar ratio $r$. 
At first order in the slow roll approximation, taking $N=60$ we have 
\begin{equation}
    n_{s}\approx 1+2\eta_{0}-6\epsilon_{0}\approx 0.965 \quad \mbox{and} \quad r\approx 16\epsilon_{0}\approx 0.002. 
\end{equation}
Both these results are compatible not only with the latest Planck data \cite{Planck 2018 results. X. Constraints on inflation} but also with those found in the Einstein frame of Higgs inflation models like \cite{Langvik} and \cite{Higgs inflation in Einstein-Cartan gravity} or scale invariant models such as \cite{Rinaldi}.

\section{Conclusions and further work}\label{Conclusions}

We perform a generic analysis of a non-minimally coupled Brans-Dicke like model with torsion
with the intention of comparing the Einstein and the Jordan frames of the model. We initially
review some work already done in the subject by explaining that this generic model possess the
capability of being fully rewritten in terms of only the traditional Levi-Civita Ricci scalar and the
contribution of the scalar field. All the torsion components of the gravitational action are then
encoded in a non-canonical kinetic term for the scalar field.
We show that the projective transformation of the connection and the extended conformal
transformation involving the metric and the torsion tensor give rise to the same action in the
Einstein framework but the latter stems from the fact that now the Ricci scalar is constructed with
a non-symmetric connection and therefore the torsion also needs to transform under the
conformal transformation. This implication leads to a range of possibilities, as mentioned in the
main text, since how torsion is transformed is a choice one must make, therefore there could be
several interesting possible options for how torsion is transformed.
Furthermore, we find out that the Jordan frame of this type of models has the appropriate stability
conditions to possibly undergo and exit an inflationary epoch. Although the unstable fixed point
exact behavior is very model dependent, we discover that the graceful exit fixed point is a generic
property of the model. This in principle could be further studied in order to explore whether this
stable fixed point could reheat the universe, either by studying the decay rate of the scalar field
around its minimum or by exploring the particle creation process.
Some light constraints on the free parameters of the model are also explored and compared with
those found in similar contexts by other authors. This in principle is not so easy to do in a generic
model because of the complicated algebraic expressions, however we show the conditions under
which our model is able to reproduce some results found by other authors under some
reasonable assumptions.
In addition, we choose a specific form of our generic scalar field functions. The scale invariant
version of our generic model was found not only to be in agreement with observations but it also
possess a, so to say, simple form of the non-canonical kinetic energy parameter $\omega_{SI}$ of eq. (\ref{BDScaleinvariantanswer}). The fact that this parameter is just a number is a clear simplification, however this may, in
general, not be true for some more complicated models.
Finally, along the paper, we focus on the comparison between the Einstein frame, which was
already studied by several authors, and the Jordan frame of the generic model. We find out that,
although mathematically these two frames are different, several results and features of the model
arise in both frames. In conclusion we hope that some of the results presented here contribute to
the ongoing discussion on how these two frames are related. 
\acknowledgments
The authors would like to thank the National Council of Science and Technology (CONACyT) for its funding and support and an anonymous referee for his/her suggestions and comments. \\

\section*{Data Availability Statement: No Data associated in the manuscript}



\begin{thebibliography}{10}

\bibitem{The Foundation of the General Theory of Relativity}
Einstein, Albert, \emph{The Foundation of the General Theory of Relativity}, \emph{Annalen Phys.} {\bf vol 49} (1916) pg 769-822.

\bibitem{Einheitliche Feldtheorie von Gravitation und Elektrizitat}
Einstein, Albert, \emph{Einheitliche Feldtheorie von Gravitation und Elektrizitat (in German)}, \emph{Sitzungsb. Preuss. Akad. Wiss.} {\bf vol 22} (1925) pg 414.

\bibitem{Auf die Riemann-Metrik und den Fern-Parallelismus gegrundete einheitliche
Feldtheorie}
Einstein, Albert, \emph{Auf die Riemann-Metrik und den Fern-Parallelismus gegr$\ddot{u}$ndete einheitliche
Feldtheorie (in German)}, \emph{Math. Ann.} {\bf vol 102} (2015) pg 1-66. 

\bibitem{Neue Moglichkeit fur eine einheitliche Theorie von Gravitation und Elektrizitat}
Einstein, Albert, \emph{Neue Moglichkeit fur eine einheitliche Theorie von Gravitation und Elektrizitat
(in German)}, \emph{Sitzungsb. Preuss. Akad. Wiss.} (1928) pg 224.

\bibitem{Riemann-Geometrie unter Aufrechterhaltung des Begriffes des Fernparallelismus}
Einstein, Albert, \emph{Riemann-Geometrie unter Aufrechterhaltung des Begriffes des Fernparallelismus
(in German)}, \emph{Sitzungsb. Preuss. Akad. Wiss.} (1928) pg 217.

\bibitem{Teleparallel gravity}
R. Aldrovandi and J.G. Pereira, \emph{Teleparallel gravity}, \emph{Fund. Theor. Phys.} {\bf vol 173} Springer,
Dordrecht, The Netherlands (2013).

\bibitem{f(T) teleparallel gravity and cosmology}
Cai, Yi-Fu and Capozziello, Salvatore and De Laurentis, Mariafelicia and Saridakis, Emmanuel N., \emph{f(T) teleparallel gravity and cosmology}, \emph{Rept. Prog. Phys.} {\bf vol 79} pg 106901 (2016).

\bibitem{Introduction to teleparallel gravities,in9
th
Mathematical Physics Meeting:
Summer School and Conference on Modern Mathematical Physics,Belgrade,Serbia,}
A. Golovnev, \emph{Introduction to teleparallel gravities, in 9th
Mathematical Physics Meeting:
Summer School and Conference on Modern Mathematical Physics, Belgrade, Serbia,} 18-23 September 2017.

\bibitem{Torsion gravity}
Hammond, R. T., \emph{Torsion gravity}, \emph{Rept. Prog. Phys.} {\bf vol 65} pg 599-649 (2002).

\bibitem{The teleparallel equivalent of general relativity}
Maluf, J. W., \emph{The teleparallel equivalent of general relativity}, \emph{Annalen Phys.} {\bf vol 525} pg 339-357 (2013).

\bibitem{Physical aspects of the space-time torsion}
Shapiro, I. L., \emph{Physical aspects of the space-time torsion}, \emph{Phys. Rept.} {\bf vol 357} pg 113 (2002).

\bibitem{Conformal symmetry, anomaly and effective action for metric-scalar gravity with torsion}
Helayel-Neto, J. A. and Penna-Firme, A. and Shapiro, I. L., \emph{Conformal symmetry, anomaly and effective action for metric-scalar gravity with torsion}, \emph{Phys. Lett. B} {\bf vol 479} pg 411-420 (2000).

\bibitem{Higgs inflation and teleparallel gravity}
Raatikainen, Sami and Rasanen, Syksy, \emph{Higgs inflation and teleparallel gravity}, \emph{JCAP} {\bf vol 12} pg 021 (2019).

\bibitem{Symmetric teleparallel general relativity}
J.M. Nester and H.-J. Yo, \emph{Symmetric teleparallel general relativity}, \emph{Chin. J. Phys.} {\bf vol 37} pg 113 (1999).

\bibitem{Palatini}
M. Ferraris, M. Francaviglia and C. Reina, \emph{Variational formulation of general relativity from
1915to 1925“Palatini’s method” discovered by Einstein in 1925}, \emph{Gen. Rel. Grav.} {\bf vol 14} pg 243 (1982).

\bibitem{Langvik}
L\r{a}ngvik, Miklos and Ojanper\"a, Juha-Matti and Raatikainen, Sami and R\"as\"anen, Syksy, \emph{Higgs inflation with the Holst and the Nieh\textendash{}Yan term}, \emph{Phys. Rev. D} {\bf vol 103} pg 083514 (2021).

\bibitem{Higgs inflation in Einstein-Cartan gravity}
Shaposhnikov, Mikhail and Shkerin, Andrey and Timiryasov, Inar and Zell, Sebastian, \emph{Higgs inflation in Einstein-Cartan gravity}, \emph{JCAP} {\bf vol 02} pg 008 (2021).

\bibitem{PARITY VIOLATION IN METRIC TORSION THEORIES OF GRAVITATION}
Hojman, R. and Mukku, C. and Sayed, W. A. \emph{Parity violation in metric torsion theories of gravitation}, \emph{Phys. Rev. D} {\bf vol 22} pg 1915-1921 (1980).

\bibitem{Gravity With Propagating Pseudoscalar Torsion}
Nelson, Philip C. \emph{Gravity With Propagating Pseudoscalar Torsion}, \emph{Phys. Lett. A} {\bf vol 79} pg 285 (1980).

\bibitem{Barberos Hamiltonian derived from a generalized Hilbert-Palatini action}
S. Holst, \emph{Barbero’s Hamiltonian derived from a generalized Hilbert-Palatini action}, \emph{Phys. Rev.
D} {\bf vol 53} (1996).

\bibitem{An Identity in Riemann-cartan Geometry}
Nieh, H. T. and Yan, M. L. \emph{An Identity in Riemann-cartan Geometry}, \emph{J. Math. Phys.} {\bf vol 23} pg 373 (1982).


\bibitem{Einstein frame or Jordan frame?}
Faraoni, V. and Gunzig, E. \emph{Einstein frame or Jordan frame?}, \emph{Int. J. Theor. Phys.} {\bf vol 38} pg 217-225 (1999).

\bibitem{Equivalence of the Einstein and Jordan frames}
Postma, Marieke and Volponi, Marco. \emph{Equivalence of the Einstein and Jordan frames}, \emph{Phys. Rev. D.} {\bf vol 90} pg 103516 (2014).

\bibitem{Cosmological viability of f(R)-gravity as an ideal fluid and its compatibility with a matter dominated phase}
Capozziello, Salvatore and Nojiri, S. and Odintsov, S. D. and Troisi, A. \emph{Cosmological viability of f(R)-gravity as an ideal fluid and its compatibility with a matter dominated phase}, \emph{Phys. Lett. B.} {\bf vol 639} pg 135-143 (2006).


\bibitem{Higgs-Dilaton Cosmology: From the Early to the Late Universe}
Garcia-Bellido, Juan and Rubio, Javier and Shaposhnikov, Mikhail and Zenhausern, Daniel. \emph{Higgs-Dilaton Cosmology: From the Early to the Late Universe}, \emph{Phys. Rev. D.} {\bf vol 84} pg 123504 (2011).

\bibitem{Dark energy: The Equation of state description versus scalar-tensor or modified gravity}
Capozziello, S. and Nojiri, S. and Odintsov, S. D. \emph{Dark energy: The Equation of state description versus scalar-tensor or modified gravity}, \emph{Phys. Lett. B.} {\bf vol 634} pg 93-100 (2006).

\bibitem{Cosmic relic torsion from inflationary cosmology}
Garcia de Andrade, L. C. \emph{Cosmic relic torsion from inflationary cosmology}, \emph{Int. J. Mod. Phys. D} {\bf vol 8} pg 725-729 (1999).

\bibitem{Cosmological inflation driven by a scalar torsion function}
Guimar\~aes, T. M. and Lima, R. de C. and Pereira, S. H. \emph{Cosmological inflation driven by a scalar torsion function}, \emph{Eur. Phys. J. C} {\bf vol 81} pg 271 (2021).

\bibitem{Cosmology with torsion: An alternative to cosmic inflation}
Pop\l{}awski, Nikodem J. \emph{Cosmology with torsion: An alternative to cosmic inflation}, \emph{Phys. Lett. B.} {\bf vol 694} pg 181-185 (2010).

\bibitem{Einstein-Cartan Cosmologies}
Medina, Sergio Bravo and Nowakowski, Marek and Batic, Davide. \emph{Einstein-Cartan Cosmologies}, \emph{Annals Phys.} {\bf vol 400} pg 64-108 (2019).


\bibitem{Status of non-Riemannian cosmology}
Puetzfeld, Dirk. \emph{Status of non-Riemannian cosmology}, \emph{New Astron. Rev.} {\bf vol 49} pg 59-64 (2005).

\bibitem{Planck 2018 results. X. Constraints on inflation}
Akrami, Y. and others, \emph{Planck 2018 results. X. Constraints on inflation}, \emph{Astron. Astrophys.} {\bf vol 641} pg A10 (2020).

\bibitem{Cold and fuzzy dark matter}
Hu, Wayne and Barkana, Rennan and Gruzinov, Andrei, \emph{Cold and fuzzy dark matter}, \emph{Phys. Rev. Lett.} {\bf vol 85} pg 1158-1161 (2000).

\bibitem{Axion Cosmology}
Marsh, David J. E. \emph{Axion Cosmology}, \emph{Phys. Rept.} {\bf vol 643} pg 1-79 (2016).

\bibitem{The Minimal model of nonbaryonic dark matter: A Singlet scalar}
Burgess, C. P. and Pospelov, Maxim and ter Veldhuis, Tonnis, \emph{The Minimal model of nonbaryonic dark matter: A Singlet scalar}, \emph{Nucl. Phys. B.} {\bf vol 619} pg 709-728 (2001).

\bibitem{Ultralight scalars as cosmological dark matter}
Hui, Lam and Ostriker, Jeremiah P. and Tremaine, Scott and Witten, Edward, \emph{Ultralight scalars as cosmological dark matter}, \emph{Phys. Rept.} {\bf vol 95} pg 043541 (2017).

\bibitem{Strong Constraints on Fuzzy Dark Matter from Ultrafaint Dwarf Galaxy Eridanus}
Marsh, David J. E. and Niemeyer, Jens C. \emph{Strong Constraints on Fuzzy Dark Matter from Ultrafaint Dwarf Galaxy Eridanus II}, \emph{Phys. Rev. Lett.} {\bf vol 123} pg 051103 (2019).

\bibitem{Dynamics of dark energy}
Copeland, Edmund J. and Sami, M. and Tsujikawa, Shinji, \emph{Dynamics of dark energy}, \emph{Int. J. Mod. Phys. D} {\bf vol 15} pg 1753-1936 (2006).

\bibitem{Quintessence, cosmic coincidence, and the cosmological constant}
Zlatev, Ivaylo and Wang, Li-Min and Steinhardt, Paul J. \emph{Quintessence, cosmic coincidence, and the cosmological constant}, \emph{Phys. Rev. Lett.} {\bf vol 82} pg 896-899 (1999).

\bibitem{Coupled quintessence}
Amendola, Luca, \emph{Coupled quintessence}, \emph{Phys. Rev. D} {\bf vol 62} pg 043511 (2000).

\bibitem{An Alternative to quintessence}
Kamenshchik, Alexander Yu, Moschella, Ugo, Pasquier, Vincent, \emph{An Alternative to quintessence}, \emph{Phys. Lett. B} {\bf vol 511} pg 265-268 (2001).

\bibitem{Cosmological Consequences of a Rolling Homogeneous Scalar Field}
Ratra, Bharat and Peebles, P. J. E. \emph{Cosmological Consequences of a Rolling Homogeneous Scalar Field}, \emph{Phys. Rev. D} {\bf vol 37} pg 3406 (1998).

\bibitem{kinflation}
Armendariz-Picon, C. and Damour, T. and Mukhanov, Viatcheslav F. \emph{k - inflation}, \emph{Phys. Lett. B} {\bf vol 458} pg 209-218 (1999).

\bibitem{Chaotic Inflation}
Linde, Andrei D. \emph{Chaotic Inflation}, \emph{Phys. Lett. B} {\bf vol 129} pg 177-181 (1983).

\bibitem{Hybrid inflation}
Linde, Andrei D. \emph{Hybrid inflation}, \emph{Phys. Rev. D} {\bf vol 49} pg 748-754 (1994).


\bibitem{Higgs}
Aad, Georges and others, \emph{Observation of a new particle in the search for the Standard Model Higgs boson with the ATLAS detector at the LHC}, \emph{Phys. Lett. B} {\bf vol 716} pg 1-29 (2012).

\bibitem{The Standard Model Higgs boson as the inflaton}
Bezrukov, Fedor L. and Shaposhnikov, Mikhail, \emph{The Standard Model Higgs boson as the inflaton}, \emph{Phys. Lett. B} {\bf vol 659} pg 703-706 (2008).

\bibitem{Jordan-Brans-Dicke Theory}
Brans, Carl H. \emph{Jordan-Brans-Dicke Theory}, \emph{Scholarpedia} {\bf vol 9} pg 31358 (2014).

\bibitem{Starobinsky} A. A. Starobinsky, in Quantum Gravity, Proceedings of the
    2nd Seminar on Quantum Gravity, Moscow, 1981(INR
    Press, Moscow, 1982), pp. 58–72; reprinted inM. A.
    Markov and P. C. West eds., Quantum Gravity(Plenum
    Press, New York, 1984), pp. 103–128; A. A. Starobinsky,
    Phys. Lett.91B, 99102 (1980).

\bibitem{ON THE RENORMALIZATION OF MODELS OF QUANTUM FIELD THEORY IN AN EXTERNAL GRAVITATIONAL FIELD WITH TORSION}
Buchbinder, I. L. and Shapiro, I. L. \emph{On the renormalization of models of quantum field theory in an external gravitational field with torsion}, \emph{Phys. Lett. B} {\bf vol 151} pg 263-266 (1985).

\bibitem{BRANS-DICKE TYPE MODELS WITH TORSION}
German, G. \emph{Brans-Dicke type models with torsion}, \emph{Phys. Rev. D} {\bf vol 32} pg 3307-3308 (1985).

\bibitem{Higgs-Dilaton inflation in Einstein-Cartan gravity}
Piani, Matteo and Rubio, Javier. \emph{Higgs-Dilaton inflation in Einstein-Cartan gravity}, \emph{JCAP} {\bf vol 05} pg 009 (2022).

\bibitem{The Meaning of Relativity}
Einstein, Albert \emph{The Meaning of Relativity}, 5th ed. (Princeton Univ. ,
Princeton, N. J.,
1955).

\bibitem{Racioppi}
Gialamas, Ioannis D. and Karam, Alexandros and Pappas, Thomas D. and Racioppi, Antonio and Spanos, Vassilis C. \emph{Scale-invariance, dynamically induced Planck scale and inflation in the Palatini formulation}, \emph{J. Phys. Conf. Ser.} {\bf vol 2105} pg 012005 (2021).

\bibitem{Dynamically induced Planck scale and inflation in the Palatini formulation}
Gialamas, Ioannis D. and Karam, Alexandros and Racioppi, Antonio, \emph{Dynamically induced Planck scale and inflation in the Palatini formulation}, \emph{JCAP.} {\bf vol 11} pg 014 (2020).

\bibitem{Inflation and reheating in scaleinvariant scalartensor gravity}
Tambalo, Giovanni and Rinaldi, Massimiliano, \emph{Inflation and reheating in scale-invariant scalar-tensor gravity}, \emph{Gen. Rel. Grav.} {\bf vol 49} pg 52 (2017).

\bibitem{Rinaldi}
Rinaldi, Massimiliano and Vanzo, Luciano, \emph{Inflation and reheating in theories with spontaneous scale invariance symmetry breaking}, \emph{Phys. Rev. D} {\bf vol 94} pg 024009 (2016).

\bibitem{Scale-independent inflation}
Ferreira, Pedro G. and Hill, Christopher T. and Noller, Johannes and Ross, Graham G. \emph{Scale-independent $R^2$ inflation}, \emph{Phys. Rev. D} {\bf vol 100} pg 123516 (2019).

\bibitem{Scale invariance dynamically induced Planck scale and inflation in the Palatini formulation}
Gialamas, Ioannis D. and Karam, Alexandros and Pappas, Thomas D. and Racioppi, Antonio and Spanos, Vassilis C. \emph{Scale-invariance, dynamically induced Planck scale and inflation in the Palatini formulation}, \emph{J. Phys. Conf. Ser.} {\bf vol 2105} pg 012005 (2021).


\bibitem{Higgs Inflation with a Gauss-Bonnet term in the Jordan Frame}
van de Bruck, Carsten and Longden, Chris, \emph{Higgs Inflation with a Gauss-Bonnet term in the Jordan Frame}, \emph{Phys. Rev. D} {\bf vol 93} pg 063519 (2016).


 \end{thebibliography}
\end{document}